\documentclass{iau}
\usepackage{graphicx}

\title[Distance determination from the PL relations] 
{Distance determination from the Cepheids and RR Lyrae period-luminosity relations}

\author[Ngeow, Gieren \& Klein]   
{Chow-Choong Ngeow$^1$,
 Wolfgang Gieren$^2$
\and Christopher Klein$^3$}

\affiliation{$^1$Graduate Institute of Astronomy, National Central University, Jhongli 32001, Taiwan \\ email: {\tt cngeow@astro.ncu.edu.tw} \\[\affilskip]
$^2$Departamento de Astronomia, Universidad de Concepcion, Casilla 160-C, Concepcion, Chile \\
$^3$Astronomy Department, University of California, Berkeley, CA 94720, USA
}

\pubyear{2014}
\volume{301}
\pagerange{1--8}
\jname{Precision Asteroseismology}
\editors{J.A. Guzik, W.J. Chaplin, G. Handler \& A. Pigulski, eds.}
\begin{document}

\maketitle

\begin{abstract}
Cepheids and RR Lyrae stars are important pulsating variable stars in distance scale work because they serve as standard candles. Cepheids follow well-defined period-luminosity (PL) relations defined for bands extending from optical to mid-infrared (MIR). On the other hand, RR Lyrae stars also exhibit PL relations in the near-infrared and MIR wavelengths. In this article, we review some of the recent developments and calibrations of  PL relations for Cepheids and RR Lyrae stars. For Cepheids, we discuss the calibration of PL relations via the Galactic and the Large Magellanic Cloud routes. For RR Lyrae stars, we summarize some recent work in developing the MIR PL relations. 

\keywords{Cepheids, RR Lyrae stars, distance scale}
\end{abstract}

\firstsection 
\section{Introduction}

Classical Cepheids and RR Lyrae stars are pulsating stars that play a vital role in the definition of the distance scale ladder\footnote{For latest version of the distance scale ladder, see {\tt http://kiaa.pku.edu.cn/$\sim$grijs/\\distanceladder.pdf}}. This is because they are standard candles in the local Universe that permit the calibration of secondary distance indicators (e.g., the peak brightness of type Ia supernovae). The ultimate goal of the distance scale ladder is to determine the Hubble constant ($H_0$) with 1\% precision and accuracy. The existence of period-luminosity (PL) relations for Cepheids (from optical to infrared wavelengths) makes distance determination using this type of variable star possible. In this article, we review some prospects of the calibration of Cepheid PL relations and their role in the recent distance scale work (Section \ref{sec2}). RR Lyrae stars also obey a PL relation in the infrared, and we review some of the recent developments of such relations in Section \ref{sec3}. 

\section{The Cepheid period-luminosity relation}\label{sec2}

\begin{figure}[h]
\begin{center}
 \includegraphics[width=3.0in]{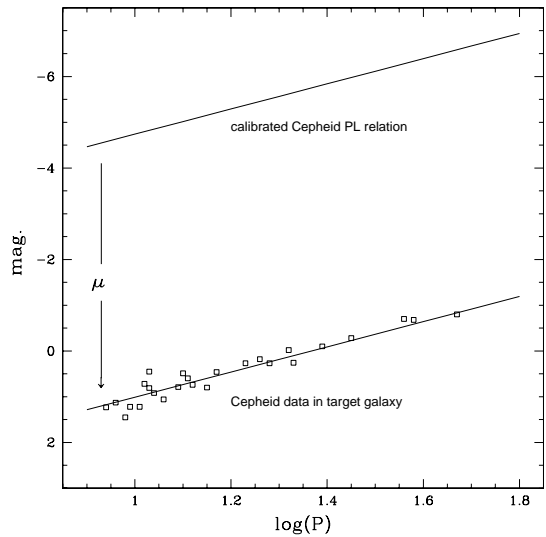} 
 \caption{Illustration of using the calibrated Cepheid PL relation to determine the distance modulus to a galaxy. After a calibrated PL relation is adopted, this calibrated PL relation is shifted vertically to fit the observed Cepheids data in a given galaxy, and the vertical offset provides the distance modulus ($\mu$) of the galaxy.}
   \label{fig0}
\end{center}
\end{figure}

The Cepheid PL relation is a 2-D projection of the period-luminosity-color (PLC) relation on the logarithmic period and magnitude plane, where the PLC relation can be derived by combining the Stefan-Boltzmann law, the period-mean density relation for pulsators, and the mass-luminosity relation based on stellar evolution models. Discussion of the physics behind the Cepheid PL relation can be found in \cite[Madore \& Freedman (1991)]{madore1991}, and will not be repeated here. The PL relation usually takes the linear form of $M_\lambda = a_\lambda \log(P) + b_\lambda$, where $a$ and $b$ are the slope and intercept of the relation in bandpass $\lambda$, respectively. Once the slopes and intercepts of the multi-band PL relations are determined or calibrated, the distance to a nearby galaxy can be obtained by fitting the calibrated PL relations to the Cepheid data in that galaxy (see Fig.~\ref{fig0}).

\subsection{Calibration of Cepheid PL relations}

Determining the slope of the PL relation is relatively straightforward. The large number of Cepheids discovered in the Magellanic Clouds permits the determination of the PL slope with $\sim$10$^{-2}$ accuracy (\cite[Soszy\'nski et al.~2008, 2010]{soszynski2008,soszynski2010}). The derivation of PL intercepts, on the other hand, is trickier, because distances to a number of Cepheids need to be known or inferred {\it a priori}. There are two routes to calibrate the Cepheid PL intercepts that are commonly found in literature: the Galactic route and the Large Magellanic Cloud route.

The Galactic route relies on Galactic Cepheids that are located in the solar neighborhood, i.e. those within few kpc. These Cepheids are bright enough that extensive data, both multi-band light curves and radial velocity curves, are available from the literature. However, they suffer from varying extinction and their distances need to be determined independently. A number of Galactic Cepheids is close enough to permit an accurate parallax measurement using {\it Hipparcos} (\cite[van Leeuwen et al.~2007]{vaL2007}) or {\it Hubble Space Telescope} ({\it HST}, \cite[Benedict et al.~2007]{benedict2007}). In the near future, {\it Gaia} will provide reliable parallaxes to almost all nearby Galactic Cepheids. Besides parallaxes, distances to Galactic Cepheids can also be determined from the Baade-Wesselink (BW) technique and its variants. The BW technique combines the measurements of radial velocities and angular diameters to derive the distance and mean radius for a given Cepheid. The angular diameter variations can be determined from the infrared surface brightness method (see, for example, \cite[Storm et al.~2011]{storm2011}, and references therein) or the interferometric technique (e.g., as in \cite[Gallenne et al.~2012]{gallene2012}). A critical parameter in the BW technique is the projection factor, or p-factor (that converts the observed radial velocity to pulsational velocity), because a 1\% error in the p-factor translates to a 1\% error in the derived distance. For a Cepheid located in an open cluster, the distance to the Cepheid can be inferred from the distance of its host cluster measured via isochrone fitting (\cite[Turner 2010]{turner2010}). Finally, the distance to a large number of Cepheids can be obtained from the calibrated Wesenheit function using {\it HST} parallaxes (\cite[Ngeow 2012]{ngeow2012}). Examples of PL relations based on Galactic Cepheids can be found in \cite[Tammann et al.~(2003)]{tammann2003}, \cite[Ngeow \& Kanbur (2004)]{ngeow2004} and \cite[Fouqu{\'e} et al.~(2007)]{fouque2007}. It has been argued that the PL relations calibrated with Galactic Cepheids are preferred in distance scale work (see \cite[Tammann et al.~2003]{tammann2003}; \cite[Kanbur et al.~2003]{kanbur2003} and reference therein), because the spiral galaxies that are used to calibrate the secondary distance indicators have metallicities close to solar value, and hence a metallicity correction to the Cepheid PL relation is not needed to derive distances in this way.

The Large Magellanic Cloud (LMC), located $\sim$50~kpc away, is an irregular galaxy that is far enough to assume that Cepheids in this galaxy lie at the same distance. Yet the LMC is also close enough that stars observed there can be resolved. Therefore, the LMC Cepheids have been commonly used in the previous studies on calibrating the Cepheid PL relations. However, measurements of the LMC distance modulus ($\mu_{\rm LMC}$) show a wide spread, ranging from $\sim$18.0 to $\sim$19.0~mag, with a center around $18.5\pm0.1$~mag (for example, see \cite[Freedman et al.~2001, Benedict et al.~2002, Schaefer 2008]{freedman2001,benedict2002,shaefer2008})\footnote{Also, see the LMC distance moduli compiled in {\tt http://clyde.as.utexas.edu/SpAstNEW/\\ head602.ps}}. This causes the calibration of the PL intercepts to suffer a systematic error of the order of $\sim$5\% (\cite[Freedman et al.~2001]{freedman2001}). For this reason, some of the PL relations derived from the LMC Cepheids leave the PL intercepts un-calibrated (i.e., the values are taken from fitting only), as shown in \cite[Soszy\'nski et al.~(2008)]{soszynski2008} and \cite[Ngeow et al.~(2009)]{ngeow2009}. Nevertheless, this problem is solved with the latest result published by \cite[Pietrzy{\'n}ski et al.~(2013)]{pietrzynski2013}. By using late-type eclipsing binary systems, they determined the distance to the LMC with 2\% accuracy, i.e., $\mu_{\rm LMC}=18.493\pm0.048$ (total error). Then, the PL relations for fundamental mode LMC Cepheids given in \cite[Soszy\'nski et al.~(2008)]{soszynski2008} become: $V=-2.762(\pm0.022)\log P - 0.963 (\pm0.015)$, $I=-2.959(\pm0.016)\log P - 1.614(\pm0.010)$ (both uncorrected for extinction), and $W=-3.314(\pm0.009)\log P -2.600 (\pm0.006)$. Similarly, the multi-band PL relations from \cite[Ngeow et al.~(2009)]{ngeow2009} can be calibrated, which is summarized in Table \ref{tab1}. 

\begin{table}[h]
  \begin{center}
    \caption{Examples of the calibrated multi-band LMC PL relations.}
    \label{tab1}
          {\scriptsize
            \begin{tabular}{|c|c|c|c|}\hline 
              {\bf Band} & {\bf Slope} & {\bf Fitted Intercept} & {\bf Calibrated Intercept}  \\ \hline
              $V$     & $-2.769\pm0.023$ & $17.115\pm0.015$ & $-1.378$ \\
              $I$     & $-2.961\pm0.015$ & $16.629\pm0.010$ & $-1.864$ \\
              $J$     & $-3.115\pm0.014$ & $16.293\pm0.009$ & $-2.200$ \\
              $H$     & $-3.206\pm0.013$ & $16.063\pm0.008$ & $-2.430$ \\
              $K$     & $-3.194\pm0.015$ & $15.996\pm0.010$ & $-2.497$ \\
              3.6 $\mu\mathrm{m}$ & $-3.253\pm0.010$ & $15.967\pm0.006$ & $-2.526$ \\
              4.5 $\mu\mathrm{m}$ & $-3.214\pm0.010$ & $15.930\pm0.006$ & $-2.563$ \\
              5.8 $\mu\mathrm{m}$ & $-3.182\pm0.020$ & $15.873\pm0.015$ & $-2.620$ \\
              8.0 $\mu\mathrm{m}$ & $-3.197\pm0.036$ & $15.879\pm0.034$ & $-2.614$ \\
              $W$     & $-3.313\pm0.008$ & $15.892\pm0.005$ & $-2.601$ \\ \hline
            \end{tabular}
          }
  \end{center}
  \vspace{1mm}
  \scriptsize{
    {\it Note:}
    The PL relations are taken from \cite[Ngeow et al.~(2009)]{ngeow2009}, calibrated with $\mu_{\rm LMC}=18.493$. Extinction corrections have been applied to the data prior to the fitting of PL relations.}
\end{table}

Two additional issues need to be taken into account when calibrating the LMC PL relations: extinction correction and non-linearity of the LMC PL relation. The LMC is known to suffer from differential extinction, hence extinction corrections need to be applied to individual LMC Cepheids by means of extinction maps (e.g., \cite[Zaritsky et al.~2004, Haschke et al.~2011]{zaritsky2004,haschke2011}). The LMC PL relation is also known to be non-linear in optical bands: the PL relation can be split into two relations separated at 10 days (for examples, see \cite[Sandage et al.~2004, Kanbur \& Ngeow 2004, Ngeow et al.~2005, Garc{\'{\i}}a-Varela et al.~2013]{sandage2004,kanbur2004,ngeow2005,garcia2013}). Both these issues, nevertheless, can be remedied by using the Wesenheit function (\cite[Madore \& Freedman 1991, Ngeow \& Kanbur 2005, Madore \& Freedman 2009, Ngeow et al.~2009, Bono et al.~2010, Inno et al.~2013]{madore1991,ngeow2005a,madore2009,ngeow2009,bono2010,inno2013}) or moving to the mid-infrared (MIR, from $\sim$3~$\mu\mathrm{m}$ to $\sim$10~$\mu\mathrm{m}$, \cite[Freedman et al.~2008, Ngeow \& Kanbur 2008, Madore et al.~2009, Ngeow et al.~2010, Scowcroft et al.~2011]{freedman2008,ngeow2008,madore2009a,ngeow2010,scowcroft2011}) at which extinction is negligible.

\subsection{Examples of distance scale application}

Both the {\it HST} $H_0$ Key Project (\cite[Freedman et al.~2001]{freedman2001}) and SN Ia {\it HST} Calibration Pro\-gram (\cite[Sandage et al.~2006]{sandage2006}), two benchmark programs that utilized the Cepheid PL relation in distance scale work, derived a Hubble constant with a $10$\% uncertainty. Since then, two additional programs, the SH0ES (Supernovae and $H_0$ for the Equation of State, \cite[Riess et al.~2011]{riess2011}) and the CHP (Carnegie Hubble Program, \cite[Freedman et al.~2012]{freedman2012}), ai\-med to determine the Hubble constant with a $3$\% uncertainty by reducing or eliminating various systematic errors. Again, the Cepheid PL relation plays an important role in these programs. One of the main differences between the SH0ES program and previous programs is that in the SH0ES program the LMC was replaced with NGC 4258 as an anchoring galaxy in the determination of the distance scale ladder. In NGC 4258, the motions of water masers surrounding its central black hole permit an accurate geometrical distance to be determined (\cite[Humphreys et al.~2008]{humphreys2008}). To further reduce the systematic errors along the distance scale ladder, the SH0ES program adopted only ``ideal'' type Ia supernovae in nearby galaxies. They are used to calibrate their peak brightness, using a homogeneous sample of Cepheids, and observed with a single instrument on-board the {\it HST}. The CHP, on the other hand, recalibrated the {\it HST} $H_0$ Key Project distance scale ladder by adopting the MIR PL relation, where the PL slopes are defined by the LMC Cepheids and the PL intercepts are calibrated with Galactic Cepheids that have {\it HST} parallaxes. Similar to SH0ES, CHP also utilized only a single instrument on-board the {\it Spitzer Space Telescope} to derive and calibrate the MIR Cepheid PL relations. Both programs derived the Hubble constant with an uncertainty of $\sim$3\%.

\section{Period-luminosity relations for RR Lyrae stars}\label{sec3}

RR Lyrae stars follow PL relations in optical to infrared bands. However, the $V$-band bolometric correction for RR Lyrae stars is almost independent of temperature, suggesting the slope of their $V$-band PL relation is zero or very close to it (instead, RR Lyrae stars follow an $M_V$-[Fe/H] relation in the $V$-band). In contrast, there is a temperature dependence of the bolometric correction in infrared bands, which translates to an observed $K$-band PL relation (\cite[Bono et al.~2001, Bono 2003]{bono2001,bono2003}). The observed $K$-band PL relation for RR Lyrae stars can be dated back to \cite[Longmore et al.~(1986)]{longmore1986}, who derived the relation based on single-epoch observations of RR Lyrae stars in three globular clusters. Recent calibration of the $K$-band PL relation, or the PL$_K$-[Fe/H] relation, can be found in, for example, \cite[Sollima et al.~(2006)]{sollima2006}, \cite[Borissova et al.~(2009)]{borissova2009}, \cite[Benedict et al.~(2011)]{benedict2011} and \cite[Dambis et al.~(2013)]{dambis2013}. When calibrating the $K$-band PL relation with RR Lyrae stars in globular clusters, one has to be cautious because RR Lyrae stars near the cluster's core may suffer from blending (\cite[Majaess et al.~2012]{majaess2012}).

\begin{figure}[h]
\begin{center}
 \includegraphics[width=3.0in]{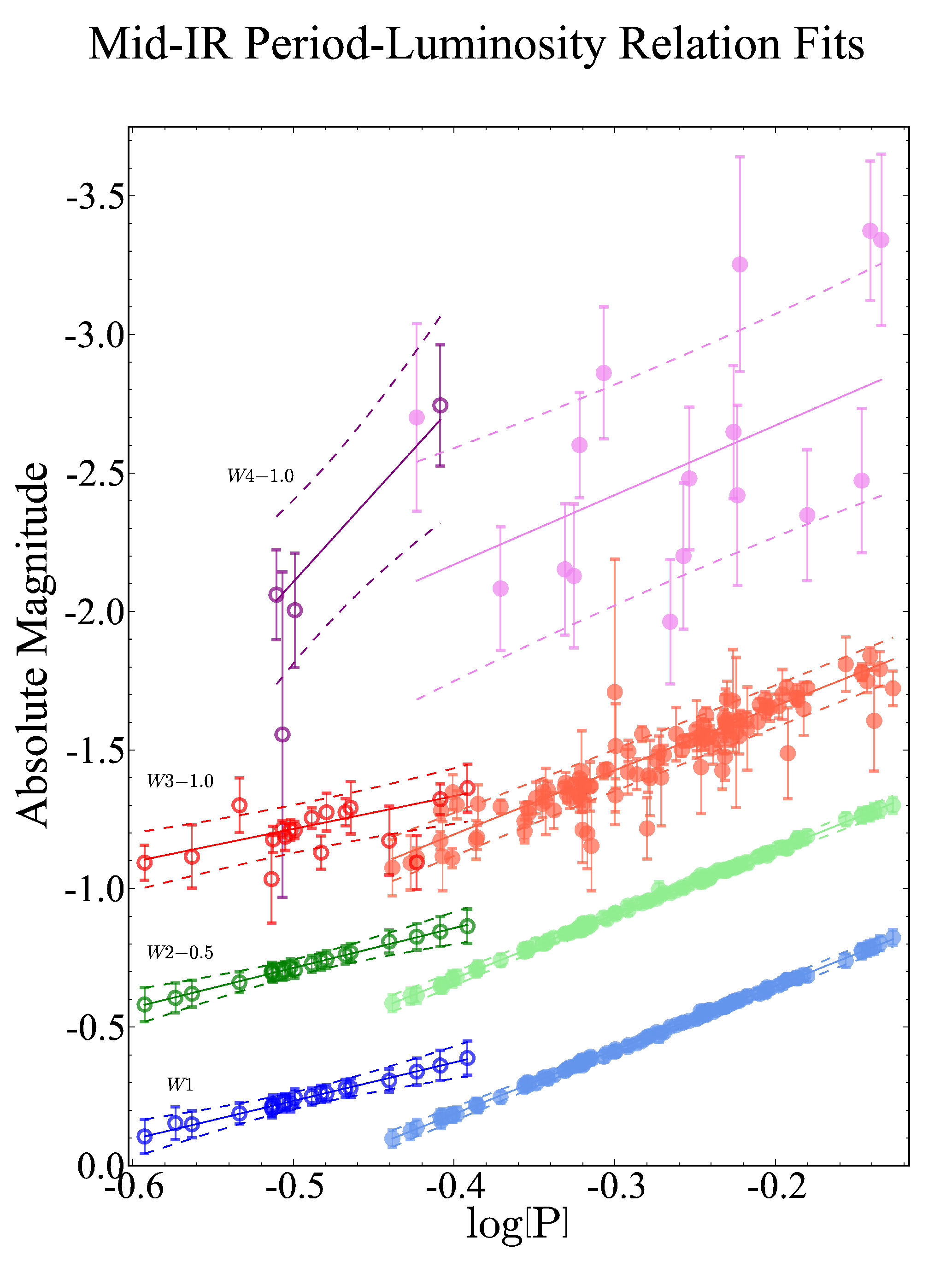} 
 \caption{Preliminary RR Lyrae stars PL relations in {\it WISE's} bands based on $143$ field RR Lyrae stars. Filled and open circles represent the RR Lyrae stars of both Bailey $ab$ and $c$ type, respectively.}
   \label{fig1}
\end{center}
\end{figure}

The derivation of the PL relation for RR Lyrae stars can be extended to MIR wavelengths. This is convincingly demonstrated by \cite[Klein et al.~(2011)]{klein2011}, who derived the MIR PL relations in {\it Wide-field Infrared Survey Explorer (WISE)} $W1$ (3.4 $\mu\mathrm{m})$, $W2$ (4.6 $\mu\mathrm{m})$ and $W3$ (12 $\mu\mathrm{m})$ bands for 76 field RR Lyrae stars. When deriving these PL relations, \cite[Klein et al.~(2011)]{klein2011} employed a Bayesian framework where the posterior distances were based on the data from {\it Hipparcos}. An updated version of the MIR PL relations with nearly double the sample size is shown in Fig. \ref{fig1}. Independently, \cite[Madore et al.~(2013)]{madore2013} derived similar MIR PL relations based on four Galactic RR Lyrae stars having parallaxes measured by the {\it HST}.

\section{Conclusion}

Independent measurements of the Hubble constant via the distance scale ladder are expected to achieve $\sim$1\% uncertainty in the future. This is possible due to a large number of Cepheids and RR Lyrae stars with high-quality data which will become available from various future or on-going projects, such as {\it Gaia}, the fourth-phase of the Optical Gravitational Lensing Experiment (OGLE-IV), and the VISTA survey of the Magellanic Clouds (VMC). The {\it James Webb Space Telescope} (\textit{JWST}), which will operate mainly in the MIR, is expected to routinely observe Cepheids beyond 30 Mpc, and it is also expected that data from this satellite will allow to derive a Hubble constant with a 1\% uncertainty. Therefore, accurate and independent calibrations of the PL relations for Cepheids and RR Lyrae stars in the MIR are important in the preparation for the {\it JWST} era.

\acknowledgements
We would like to thank the invitation of the SOC and the LOC for presenting this talk at the conference. CCN acknowledges the support from NSC grant NSC101-2119-M-008-007-MY3.

\end{document}